\documentclass[letterpaper]{article} % DO NOT CHANGE THIS
\usepackage{aaai25}  % DO NOT CHANGE THIS
\usepackage{times}  % DO NOT CHANGE THIS
\usepackage{helvet}  % DO NOT CHANGE THIS
\usepackage{courier}  % DO NOT CHANGE THIS
\usepackage[hyphens]{url}  % DO NOT CHANGE THIS
\usepackage{graphicx} % DO NOT CHANGE THIS
\usepackage{natbib}  % DO NOT CHANGE THIS AND DO NOT ADD ANY OPTIONS TO IT
\usepackage{caption} % DO NOT CHANGE THIS AND DO NOT ADD ANY OPTIONS TO IT
\usepackage{algorithm}
\usepackage{algorithmic}
\usepackage{amssymb}
\usepackage{amsmath}
\usepackage{multicol}
\usepackage{multirow}
\usepackage{booktabs}

\usepackage{newfloat}
\usepackage{listings}
\DeclareCaptionStyle{ruled}{labelfont=normalfont,labelsep=colon,strut=off} % DO NOT CHANGE THIS
\floatstyle{ruled}
\newfloat{listing}{tb}{lst}{}
\floatname{listing}{Listing}
\title{SegX: Improving Interpretability of Clinical Image Diagnosis with Segmentation-based Enhancement}
\author {
    Yuhao Zhang\textsuperscript{\rm 1, \equalcontrib},
    Mingcheng Zhu\textsuperscript{\rm 2, \equalcontrib},
    Zhiyao Luo\textsuperscript{\rm 2, \thanks{Corresponding author: zhiyao.luo@eng.ox.ac.uk}}
}
\affiliations {
    \textsuperscript{\rm 1}National University of Singapore, Singapore\\
    \textsuperscript{\rm 2}Department of Engineering Science, University of Oxford, UK\\
    yuhao\_zhang@u.nus.edu, mingcheng.zhu@eng.ox.ac.uk, zhiyao.luo@eng.ox.ac.uk
}

\usepackage{bibentry}
\begin{document}

\maketitle

\begin{abstract}
Deep learning-based medical image analysis faces a significant barrier due to the lack of interpretability. Conventional explainable AI (XAI) techniques, such as Grad-CAM and SHAP, often highlight regions outside clinical interests. To address this issue, we propose Segmentation-based Explanation (SegX), a plug-and-play approach that enhances interpretability by aligning the model's explanation map with clinically relevant areas leveraging the power of segmentation models. Furthermore, we introduce Segmentation-based Uncertainty Assessment (SegU), a method to quantify the uncertainty of the prediction model by measuring the 'distance' between interpretation maps and clinically significant regions. Our experiments on dermoscopic and chest X-ray datasets show that SegX improves interpretability consistently across mortalities, and the certainty score provided by SegU reliably reflects the correctness of the model's predictions. Our approach offers a model-agnostic enhancement to medical image diagnosis towards reliable and interpretable AI in clinical decision-making.
\end{abstract}

% \keywords{Explainable AI, Medical Image Analysis, Segmentation-based Explainability, Uncertainty Assessment}

\begin{links}
    \link{Code}{https://github.com/JasonZuu/SegX}
    % \link{Datasets}{https://www.kaggle.com/datasets/mingchengzhu/segx-used-datasets}
\end{links}

% \begin{quote}\begin{scriptsize}\begin{verbatim}
% \documentclass[letterpaper]{article}
% \usepackage[submission]{aaai25}
% \end{verbatim}\end{scriptsize}\end{quote}

\section{Introduction}
The integration of artificial intelligence (AI) into medical image analysis has become a central area of research due to its potential to streamline workflows and enhance diagnostic accuracy \cite{pinto2023artificial}. Among various AI techniques, deep learning (DL) have shown remarkable efficacy across diverse medical applications, including disease diagnosis and lesion segmentation \cite{barata2023reinforcement, gu2018recent, ramesh2021review}. For instance, on the HAM10000 dataset \cite{tschandl2018ham10000}, a diagnostic model has achieved a high sensitivity of 87.1\% for basal cell carcinoma detection based on reinforcement learning \cite{barata2023reinforcement}, which even outperforming the clinicians. Similarly, a model that learns the common patterns from multi-source data achieved an accuracy of 85.8\% across multiple diseases on the ChestX-Det10 dataset \cite{liu2020chestx}, demonstrating DL's potential in accurate chest X-ray diagnosis \cite{dai2024unichest}. Despite these advancements, the black-box nature of these models and their opaque decision-making processes remain critical barriers to their adaptation in broader clinical scenarios, where interpretability and trustworthiness are paramount \cite{abbasian2024interpretation}.

\begin{figure}[ht!]
    \centering
    \includegraphics[width=1.0\linewidth]{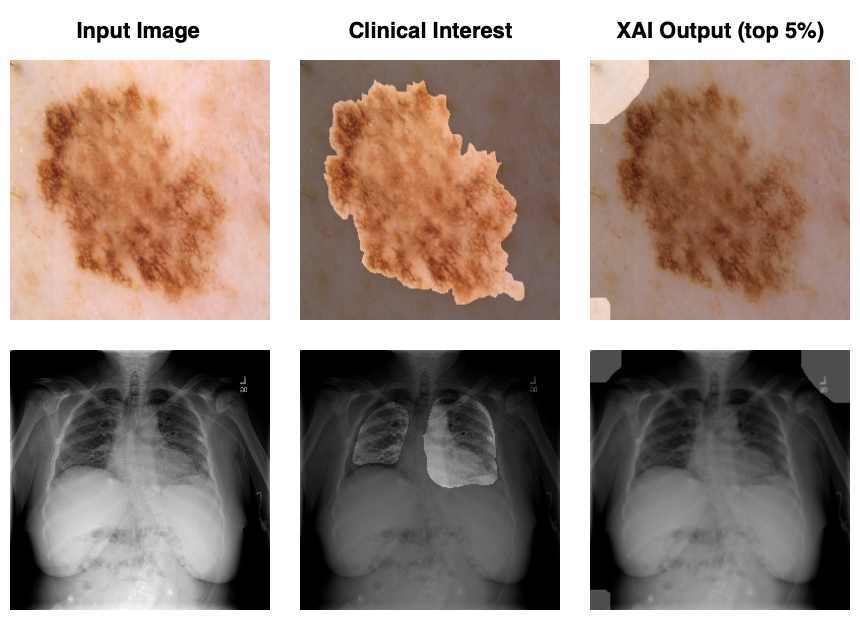}
    \caption{\textbf{Negative examples of using XAI to highlight highly important regions.} The first column shows the original images from HAM10000 and ChestX-Det10, respectively. The second column shows the clinical interests of each image. The last column shows the Grad-CAM interpreted results on a learned model. It is clear to see that the model overemphasises background area for making a diagnosis, which mismatches clinical interests.}
    \label{fig:XAI problem}
\end{figure}

Explainable AI (XAI) methods have demonstrated significant value in enhancing model interpretability across various computer vision tasks by providing visual evidence for predictions, which is often intuitive for human users \cite{hassija2024interpreting}. For instance, Grad-CAM \cite{selvaraju2017grad} highlights regions in an image that contribute most to the model's prediction by leveraging the gradients of target class scores with respect to feature map activations. In medical image classification, methods like Grad-CAM \cite{selvaraju2017grad} and SHAP \cite{lundberg2017unified} have been widely adopted to enhance the trustworthiness of AI models \cite{loh2022application, borys2023explainable}. However, as shown in Fig.\ref{fig:XAI problem}, these methods fail to consistently align their explanation regions with clinically relevant areas \cite{yuan2024clinical}, which limits their applicability in clinical practice. 

Knowledge-guided XAI (KG-XAI) methods have been developed to enhance the clinical relevance of model explanations by integrating clinical knowledge into the interpretability process \cite{yuan2024clinical, jung2023weakly, yuan2023leveraging}. These methods extract clinically significant regions from input images using techniques such as image transformations \cite{yuan2024clinical} or deep learning-based approaches \cite{jung2023weakly, yuan2023leveraging}, which serve as clinical knowledge. The extracted clinical knowledge is then used to align XAI outputs with these regions, resulting in explanation maps that better reflect clinical interest. However, KG-XAI methods overlook regions outside the identified clinical areas, such as background features, which can also carry meaningful information. For example, the background may contain contextual cues or subtle signs of disease progression \cite{cozzi2021ground}. masking background regions outside the clinical areas can lead to information loss and potentially mislead clinicians by providing an incomplete understanding of the model's predictions.

\begin{figure*}[h]
    \centering
    \includegraphics[width=0.9\linewidth]{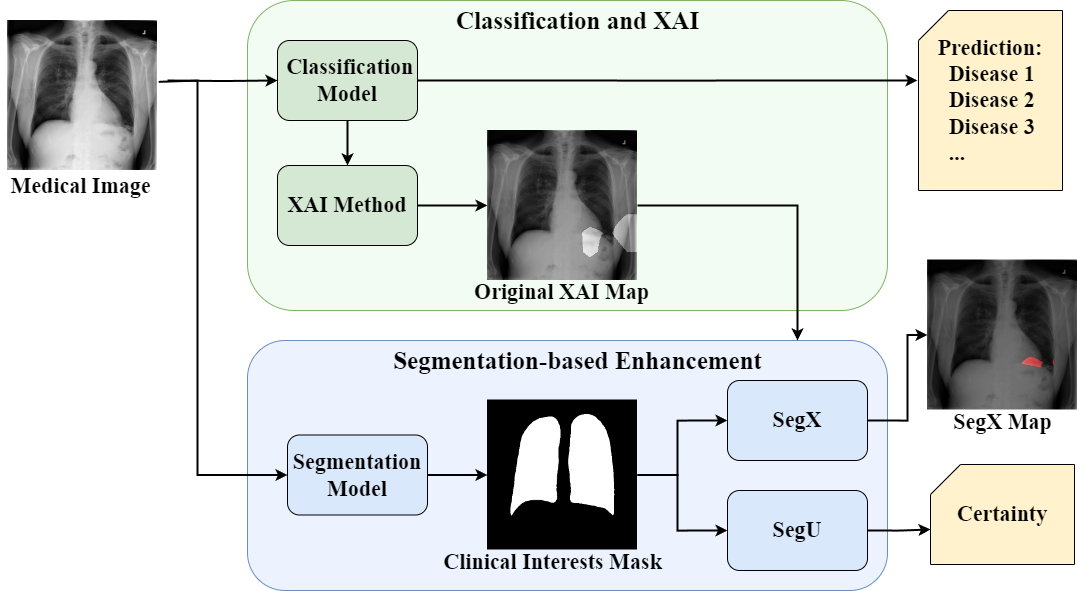}
    \caption{\textbf{Overview of the proposed method for improving XAI outputs with segmentation-based improvement.} The process begins with the classification and XAI pipeline (green), where a medical image is processed by a classification model to generate predictions and an XAI method produces the original explanation map. Afterwards, the segmentation-based enhancement (blue) starts with a segmentation model that generates a clinical interests mask to align the XAI outputs with clinically significant regions using the SegX module, producing a refined SegX map. Additionally, the SegU module quantifies the certainty of the prediction by evaluating the alignment between the explanation map and the clinical interests mask. Together, SegX and SegU enhance the interpretability and reliability of the original process, supporting better decision-making in clinical applications.}
    \label{fig:overall framework}
\end{figure*}

To address the limitations of existing KG-XAI, this paper introduces Segmentation-based explanation (SegX) and Segmentation-based Uncertainty Assessment (SegU), two novel approaches that integrate clinical knowledge of disease-specific focus regions to enhance the reliability and interpretability of existing XAI techniques, as shown in Fig~\ref{fig:overall framework}.  Specifically, SegX applies the segmentation masks of clinically significant regions on the XAI output to generate refined explanation maps. SegU calculates the 'distance' between the segmentation mask and the original explanation map, therefore measuring uncertainty. 

\textbf{Designed as flexible, plug-and-play modules, SegX and SegU can be seamlessly integrated into any XAI method for medical image classification models to incorporate clinical knowledge.} Experiments on dermoscopic and X-ray datasets demonstrate that SegX consistently produces refined explanations that closely match clinician annotations. In addition, we found that SegU can distinguish between the correct and incorrect predictions of the classification model. Overall, The enhanced interpretability and uncertainty assessment provide clinicians with a better understanding of AI predictions and improve confidence levels, enabling more informed and cautious clinical decisions \cite{zhang2020inprem}, thereby facilitating the adoption of AI in clinical practice.

\section{Related Work} 
\subsection{Explanable AI (XAI)}
XAI approaches for image analysis models are designed to enhance transparency by highlighting regions within the input data that contribute significantly to the model's decision-making process. Among these approaches, class activation maps (CAM) \cite{zhou2016learning} utilise global average pooling (GAP) within CNNs to identify regions within the input associated with specific classes, enabling the model to highlight the areas most relevant for classification. Grad-CAM \cite{selvaraju2017grad} builds upon CAM by backpropagating gradients to specified layers, generating a heatmap that visualises the most influential regions in the input image; this approach is widely used in tasks such as disease diagnosis for its ability to focus on high-importance areas without altering the network architecture. SHAP \cite{lundberg2017unified}, a feature attribution method based on Shapley values, assigns importance scores to individual features, capturing the contribution of each pixel or region in complex, non-linear networks; this technique provides a more comprehensive view of feature influence across the network but requires substantial computational resources. While CAM and Grad-CAM primarily highlight relevant regions, SHAP offers pixel-level feature attributions.

Although CNN-based XAI methods such as CAM, Grad-CAM, and SHAP have shown effectiveness in conventional computer vision tasks, their explanations often lack alignment with clinically meaningful regions in medical images \cite{yuan2024clinical}. These methods may highlight irrelevant background regions due to learning and/or data biases rather than true clinically relevant regions, limiting their utility in practice.

\subsection{Knowledge-guided XAI}
Recent research focuses on aligning XAI outputs with clinical interests to enhance their relevance for clinical decision-making. These methods can be broadly categorised based on how they model clinical knowledge. One type involves extracting clinically significant regions through image transformation. For instance, \citeauthor{yuan2024clinical} \cite{yuan2024clinical} proposed a clinical template for pneumothorax by employing image transformation techniques such as flipping and dilation. This template is then applied post hoc to refine XAI outputs. While this template-based method improves the alignment effectively for the pneumothorax in chest X-ray images, its static design may limit its generalisation across different diseases and imaging modalities.

The second type of clinical knowledge modelling leverages deep learning (DL) methods to extract clinically significant regions. For instance, \citeauthor{jung2023weakly} \cite{jung2023weakly} introduced a spatial attention mechanism that identifies potential disease areas through learning, with the attention map then used to refine CAM. However, this method tightly couples the attention mechanism with the prediction process, resulting in a high dependency between clinical knowledge modelling and the model’s predictions. Consequently, it fails to provide precise explanations for incorrect predictions, limiting its interpretability. In contrast, \citeauthor{yuan2023leveraging} \cite{yuan2023leveraging} proposed a segmentation-based approach that decouples knowledge modelling from disease diagnosis. Their method employs segmentation models to create clinical masks directly from original images, refining XAI outputs independently of the prediction process. However, this approach is task-specific, as the segmentation model is designed for pneumothorax and lacks the flexibility to generalise to multi-disease scenarios. Additionally, by focusing on a narrow region only for pneumothorax, the method may overlook other relevant areas, potentially leading to significant information loss and limiting the comprehensiveness of the explanation.

While these methods improve alignment between XAI outputs and clinically significant areas, they have notable drawbacks. By focusing solely on clinically significant regions, they often mask out surrounding areas, including background regions, which can still carry important contextual information. Background features may provide subtle cues or additional context for understanding the model’s predictions. Masking these regions entirely can lead to information loss and may mislead clinicians by providing an incomplete view of the model’s decision-making process, potentially causing them to overestimate the model's reliability.

\section{Methodology}
\subsection{Problem Formulation}
\paragraph{Disease Diagnosis from Medical Images.}
Given an input image $ \mathbf{x}_i \in \mathbb{R}^{W\times H\times C}$ with width $W$, height $H$, and channel $C$, the classification model $ f_{\text{cls}} $ generates a probability distribution over multiple labels, represented as

\begin{equation}
\mathbf{p}_i = f_{\text{cls}}(\mathbf{x}_i),
\end{equation}
where $ \mathbf{p}_i = \{p_{ij} \mid j = 1, \ldots, N\} $ is the set of probability scores for each of $ N $ classes. Each element $ p_{ij} $ represents the probability of class $ j $ for the input image $ \mathbf{x}_i $.

The multi-label prediction $ \widehat{\mathbf{y}}_{i} $ is determined by a class-wise threshold $ \tau_j $. Specifically,
\begin{equation}
\widehat{\mathbf{y}}_{i} = \{ j \mid p_{ij} > \tau_j \}.
\end{equation}

The objective of the classification model is to minimise the overall prediction error across the dataset by optimizing the model parameters $ \theta $. Specifically, the objective function $\mathcal{O}_{\text{cls}}$is defined as:

\begin{equation}
\mathcal{O}_{\text{cls}} = \arg\min_\theta \frac{1}{N} \sum_{i=1}^N \mathcal{E}(\widehat{\mathbf{y}}_i, \mathbf{y}_i),
\end{equation}
where $N$ is the number of samples, $\mathbf{y}_i$ is the corresponding ground truth label set, and $ \mathcal{E} $ represents a general error metric that quantifies the difference between predictions and ground truth.

\paragraph{Explanable AI (XAI).}
XAI method $ f_{\text{xai}}$ is applied to the classification model to produce the interpretability mask $ \mathbf{m}_{i,j}\in \mathbb{R}^{H\times W} $ for each label $ j \in \widehat{\mathbf{y}}_{i}$.
\begin{equation}
\mathbf{m}^{\text{xai}}_{i,j} = f_{\text{xai}}(f_{\text{cls}}, X_i, j),
\end{equation}

 $\mathbf{m}_{i,j}$ highlights the regions in $x_i$ that makes the model $f_{\text{cls}}$ to output $p_{i,j}$.

\paragraph{Segmentation-guided XAI.}
A segmentation model \( f_{\text{seg}} \) takes the input image \( \mathbf{x}_i \) to generate a segmentation mask \( \mathbf{m}^{\text{seg}}_{i}\in \mathbb{R}^{H \times W} \) of the clinically relevant regions.
\begin{equation}
\mathbf{m}^{\text{seg}}_{i} = f_{\text{seg}}(X_i),
\end{equation}

For each label \( j \in \widehat{\mathbf{y}}_{i} \), a segmentation-based interpretability mask is generated by overlaying the interpretability mask \( \mathbf{m}^{\text{xai}}_{i,j} \) with the segmentation mask \( \mathbf{m}^{\text{seg}}(\mathbf{x}_i) \):
\begin{equation}
\mathbf{m}^{\text{segX}}_{i,j} = \mathbf{m}^{\text{xai}}_{i,j} \cap \mathbf{m}^{\text{seg}}_{i}.
\end{equation}

\subsection{Disease Diagnosis and XAI}

This study implemented CNN models \cite{he2016deep, huang2017densely} for disease diagnosis from different types of medical images, including dermoscopic images \cite{tschandl2018ham10000} and X-ray images \cite{liu2020chestx}. For an input image \( \mathbf{x} \), the CNN-based model scans and analyses the image to extract disease-related features. These features are then processed to generate a probability distribution over potential diseases.

To interpret the model's prediction, we employ XAI approaches \cite{selvaraju2017grad, lundberg2017unified} to identify regions contributing to predictions. Specifically, the XAI method calculates the influence of each feature map on the model’s output by weighting feature activations according to their gradient values. This weighted sum of feature maps is processed through an activation function \( f_{\text{act}}(\cdot) \), producing interpretability output. Following previous studies \cite{yuan2023leveraging, yuan2024clinical}, we extract the top 5\% saliency map \( \mathbf{m}^{\text{xai}}\) as the explanation map from the XAI output. This map reveals spatial and feature-specific insights, offering a clear understanding of the model's decision-making process.

\subsection{Segmentation-based Explanation (SegX)}
The SegX approach enhances interpretability by aligning the XAI's output for the classification model with clinically relevant regions yielded by a segmentation model. By constraining the XAI output to the critical clinical regions, this approach provides a more detailed and reliable explanation of the CNN model’s decision-making process, making the interpretability output more meaningful and useful for clinical applications.

Specifically, given an input image \( \mathbf{x}_i \), the segmentation model generates a mask \( \mathbf{m}^{\text{seg}}_{i,j} \), where each pixel \(\mathbf{m}^{\text{seg}}_{i,j} \in \{0,1\} \). The segmentation mask \( \mathbf{m}^{\text{seg}}_{i,j} \) indicates the region of clinical interest for the specific disease $j$. To enhance the interpretability of XAI, we overlaid the segmentation mask \( \mathbf{m}^{\text{seg}}_{i,j} \) with the original XAI's saliency map \( m^{\text{xai}}_{i,j} \) for the classification model to create the segmentation-guided XAI mask \(  m^{\text{segX}}_{i,j} \):

\begin{equation}
m^{\text{segX}}_{i,j} =m^{\text{seg}}_{i,j} \cdot m^{\text{xai}}_{i,j}.
\end{equation}

As the $m^{\text{xai}}_{i,j}$ provides the classification model's focus and $m^{\text{seg}}_{i,j}$ indicates the clinical regions, the overlap mask $m^{\text{segX}}_{i,j}$ is able to highlight regions where the classification model’s focus aligns with clinically critical regions.

Since the segmentation-guided XAI masks rely on the output of the segmentation model, it is essential for the segmentation model to accurately capture clinically relevant regions. To optimise segmentation performance and ensure alignment with true lesion areas, we employ a composite loss function:
\begin{equation}
\mathcal{L}_{\text{seg}} = \lambda \mathcal{L}_{\text{CE}} + (1 - \lambda) \mathcal{L}_{\text{Dice}},
\end{equation}
where \( \lambda \) controls the balance between cross-entropy loss \cite{zhu2022recognizing} and Dice loss \cite{kato2024adaptive}. By combining these two losses, we ensure that the model achieves both pixel-level precision and accurate clinical region localisation, which are critical for reliable XAI explanations \cite{yeung2022unified}.

\subsection{Segmentation-based Uncertainty Assessment (SegU)}
To address the potential issue that knowledge-guided XAI may inadvertently obscure the model's focus on clinically irrelevant regions, we propose a segmentation-based uncertainty assessment to alert clinicians to the model's uncertain decisions. We hypothesised that the correction predictions for biomedical images are more likely to be made from the clinically significant regions. Thus, we use the Intersection over Union (IoU) between the classification model’s focus mask \( M_{\text{cls}}(X_i) \) and the segmentation mask \( M_{\text{seg}}(X_i) \) as an uncertainty metric. The segmentation mask, when trained appropriately, is hypothesised to serve as a clinically meaningful reference. A low IoU score would indicate insufficient alignment between the classification model’s attention and clinically important regions, suggesting that the interpretability output may be unreliable.

We propose two novel certainty scores to assess the model's certainty for its prediction based on the alignment between the XAI maps and clinically significant segmentation masks. The first is the \textbf{IoU-based certainty score ($c_{\text{IoU}}$)}, which quantifies the overlap between the XAI map $m^{\text{xai}}_{i,j}$ and the corresponding segmentation mask $m^{\text{xai}}_{i,j}$. It is calculated as follows:
\begin{equation}
c_{\text{IoU}} = \frac{\left|m^{\text{xai}}_{i,j} \cap m^{\text{seg}}_{i,j} \right|}{\left|m^{\text{xai}}_{i,j} \cup m^{\text{seg}}_{i,j} \right|}.
\end{equation}

The second is based on the \textbf{Area Under the IoU-Threshold Curve (AUITC)}, which provides a comprehensive measure of alignment by evaluating the IoU scores across varying thresholds used to generate the explanation map. The AUITC-based certainty score $c_\text{AUITC}$ reflects the overall consistency between the XAI outputs and the segmentation mask across different saliency levels. Formally, it is defined as:

\begin{equation}
c_\text{AUITC} = \int_0^1 c_{\text{IoU}}(\tau) \, d\tau,
\end{equation}
where $\tau$ denotes the threshold applied to the XAI output to generate the explanation map $m^{\text{xai}}$, and $c_{\text{IoU}}(\tau)$ is the IoU score calculated at a given threshold $\tau$.

\section{Experiment}

\begin{figure*}[h]
    \centering
    \includegraphics[width=0.9\linewidth]{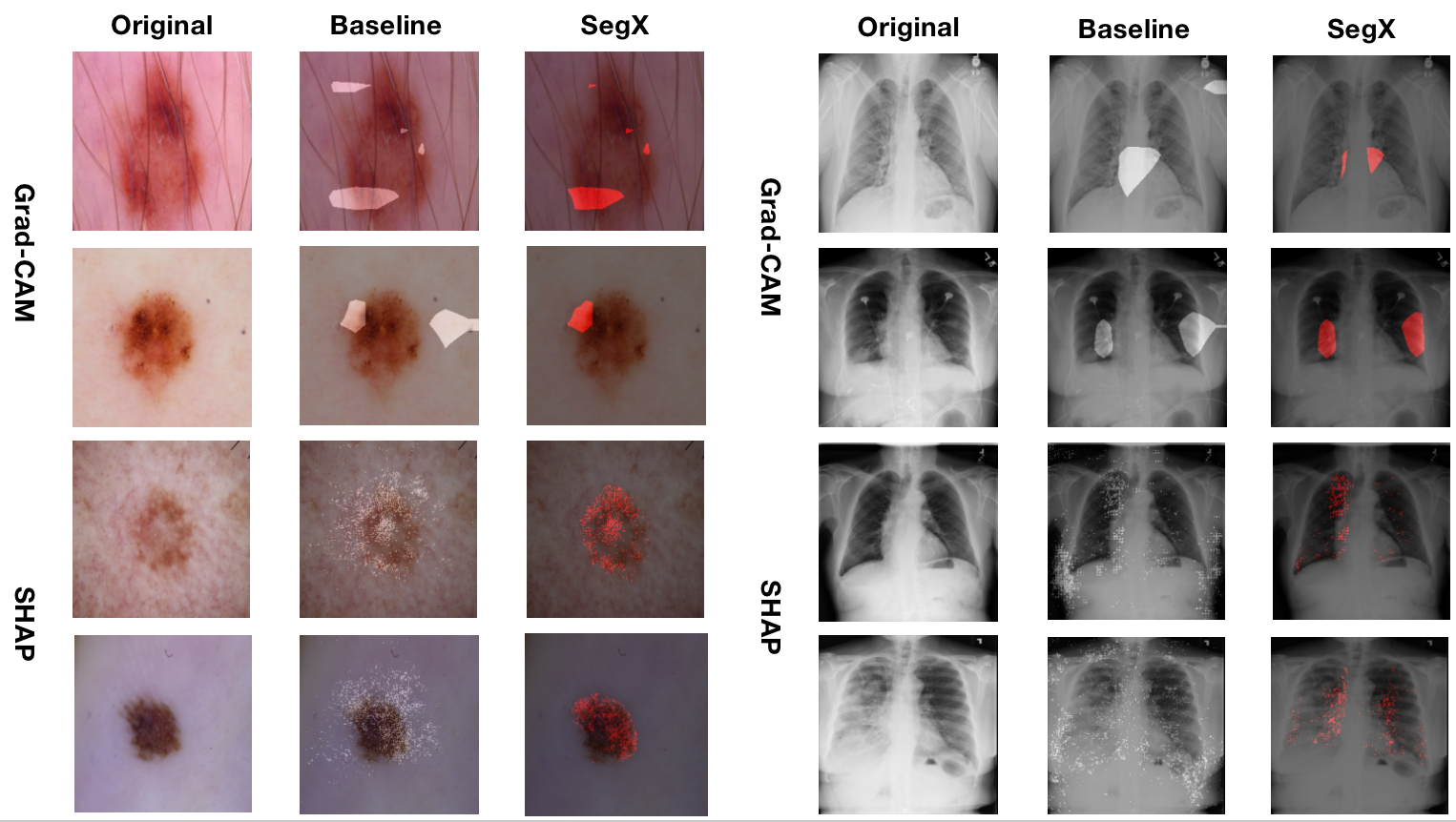}
    \caption{\textbf{Examples of original XAI masks (baseline) versus SegX masks on the HAM10000 and ChestX-Det10 test sets.} The model's top 5\% focus region is highlighted in white for the baseline and in red for the SegX method.}
    \label{fig:visual comparisons}
\end{figure*}

\subsection{Datasets}
To evaluate the effectiveness of our proposed SegX and SegU across various types of medical images, we assessed its performance using two real-world biomedical image datasets: HAM10000 \cite{tschandl2018ham10000}, a dermoscopic image dataset; and ChestX-Det10 \cite{liu2020chestx}, a chest X-ray dataset.

The HAM10000 \cite{tschandl2018ham10000} dataset is a publicly available collection of 10,015 dermoscopic images for skin lesion classification. Collected over 20 years from clinical centres in Vienna, Austria, and Queensland, Australia, this dataset contains images of seven distinct skin lesion types, encompassing both benign and malignant clinical categories. The statistics of the HAM10000 dataset are shown in Table~\ref{table:statistics of HAM10000}.

\begin{table}[h!]
\centering
\caption{Statistics for the HAM10000 Dataset}
\label{table:statistics of HAM10000}
\vspace{-5px}
\begin{tabular}{p{5cm} c}
\hline
\addlinespace[1pt]
\textbf{Lesion Type} & \textbf{Count} \\
\hline
\addlinespace[1pt]
Actinic Keratoses and Intraepithelial Carcinoma (akiec) & 327 \\
Basal Cell Carcinoma (bcc) & 514 \\
Benign Keratosis-like Lesions (bkl) & 1,099 \\
Dermatofibroma (df) & 115 \\
Melanoma (mel) & 1,113 \\
Melanocytic Nevi (nv) & 6,705 \\
Vascular Lesions (vasc) & 142 \\
\hline
\addlinespace[1pt]
Total Images & 10,015 \\
\hline
\end{tabular}
\end{table}

ChestX-Det10 \cite{liu2020chestx} is an X-ray image dataset designed for chest disease classification. It consists of 3,583 chest X-ray images annotated with labels for 10 common chest diseases or abnormalities. As each image can contain multiple co-occurring diseases, the dataset is structured for multi-label classification. The detailed statistics of ChestX-Det10 are shown in Table~\ref{table:statistics of Chestx-Det10}.

\begin{table}[h!]
\centering
\caption{Statistics for the ChestX-Det10 Dataset}
\vspace{-5px}
\label{table:statistics of Chestx-Det10}
\begin{tabular}{p{5cm} c}
\hline
\addlinespace[1pt]
\textbf{Disease} & \textbf{Count} \\
\hline
\addlinespace[1pt]
Atelectasis & 340 \\
Calcification & 347 \\
Consolidation & 2,537 \\
Effusion & 2,092 \\
Emphysema & 298 \\
Fibrosis & 738 \\
Fracture & 661 \\
Mass & 160 \\
Nodule & 955 \\
Pneumothorax & 211 \\
\hline
\addlinespace[1pt]
Total Image & 3,583 \\
\hline
\end{tabular}
\end{table}

\begin{table*}[!ht]
\centering
\caption{Comparison between XAI and SegX on HAM10000 and ChestX-Det10.}
\label{table:SegX}
\vspace{-5px}
\begin{tabular}{c c c c c c c}
\hline
\addlinespace[1pt]
\multirow{2}{*}{\textbf{Dataset}} & \multirow{2}{*}{\textbf{Model}} & \multirow{2}{*}{\textbf{XAI}} & \multicolumn{2}{c}{\textbf{IoU@5\% $\uparrow$}} & \multicolumn{2}{c}{\textbf{AUITC $\uparrow$}} \\
& & & \textbf{Original} & \textbf{SegX} & \textbf{Original} & \textbf{SegX} \\
\addlinespace[1pt]
\hline
\addlinespace[1pt]
\multirow{4}{*}{HAM10000} 
& \multirow{2}{*}{ResNet} & SHAP & 0.125 & \textbf{0.136} & 0.128 & \textbf{0.153} \\
& & Grad-CAM & 0.084 & \textbf{0.093} & 0.088 & \textbf{0.109} \\
& \multirow{2}{*}{Densenet} & SHAP & 0.121 & \textbf{0.131} & 0.128 & \textbf{0.147} \\
& & Grad-CAM & 0.176 & \textbf{0.188} & 0.178 & \textbf{0.214} \\
\hline
\addlinespace[1pt]
\multirow{4}{*}{ChestX-Det10} 
& \multirow{2}{*}{ResNet} & SHAP & 0.076 & \textbf{0.084} & 0.076 & \textbf{0.084} \\
& & Grad-CAM & 0.033 & \textbf{0.035} & 0.036 & \textbf{0.039} \\
& \multirow{2}{*}{Densenet} & SHAP & 0.061 & \textbf{0.068} & 0.061 & \textbf{0.070} \\
& & Grad-CAM & 0.052 & \textbf{0.055} & 0.054 & \textbf{0.058} \\
\hline
\end{tabular}
\end{table*}

\subsection{Experiment Settings}
We randomly divided the HAM10000 and ChestX-Det10 datasets into training, validation, and test sets in the ratio of 70:10:20, respectively. 

In this experiment, we implement a range of models and methods to evaluate the effectiveness and the generalisation of our proposed method:
\begin{itemize}
    \item Classification models: two CNN architectures were used for the classification task, including ResNet \cite{he2016deep} and DenseNet \cite{huang2017densely}, which have been proven excellent in image classification tasks and are widely used on biomedical images \cite{mall2023comprehensive}.
    \item XAI methods: To provide interpretability of the classification models, we applied two common interpretability methods to the classification model, including SHAP \cite{lundberg2017unified} and Grad-CAM \cite{selvaraju2017grad}.
    \item Segmentation model: a UNet \cite{ronneberger2015u} is used to segment clinical regions.
\end{itemize}

\begin{table*}[ht!]
\centering
\caption{Evaluation of SegU on correct and incorrect prediction groups}
\vspace{-5px}
\label{table:SegU}
\begin{tabular}{c c c c c c c}
\hline
\multirow{2}{*}{\textbf{Dataset}} & \multirow{2}{*}{\textbf{Model}} & \multirow{2}{*}{\textbf{XAI}} & \multicolumn{2}{c}{$c_\text{IoU}$$\uparrow$} & \multicolumn{2}{c}{{$c_\text{AUITC}$}$\uparrow$} \\
& & & \textbf{Correct} &\textbf{Incorrect} & \textbf{Correct} & \textbf{Incorrect}\\
\addlinespace[1pt]
\hline
\addlinespace[1pt]
\multirow{4}{*}{HAM10000} & \multirow{2}{*}{ResNet} & SHAP & \textbf{0.133} & 0.122 & \textbf{0.135} & 0.122\\
& & Grad-CAM & \textbf{0.112} & 0.015 & \textbf{0.142} & 0.022 \\
& \multirow{2}{*}{Densenet} & SHAP & \textbf{0.124 }& 0.123 & \textbf{0.124} & 0.112\\
& & Grad-CAM & \textbf{0.183} & 0.092 & \textbf{0.183} & 0.093\\
\hline
\addlinespace[1pt]
\multirow{4}{*}{ChestX-Det10} & \multirow{2}{*}{ResNet} & SHAP & \textbf{0.055} & 0.010 & \textbf{0.079} & 0.071\\
& & Grad-CAM & \textbf{0.079} & 0.072 & \textbf{0.061} & 0.014\\
& \multirow{2}{*}{Densenet} & SHAP & \textbf{0.072} & 0.055 & \textbf{0.073} & 0.055\\
& & Grad-CAM & \textbf{0.086} & 0.025 & \textbf{0.086} & 0.028\\
\hline
\end{tabular}
\end{table*}

\subsection{Results of SegX's Effectiveness Evaluation}

To evaluate the effectiveness of SegX in improving the alignment between the explanation map and the clinically significant regions, we use two metrics: the IoU and the AUITC. The IoU metric measures the overlap between two masks, defined as:

\begin{align}
\text{IoU}(\mathbf{m}_1, \mathbf{m}_2) = \frac{\left|\mathbf{m}_1 \cap \mathbf{m}_2\right|}{\left|\mathbf{m}_1 \cup \mathbf{m}_2\right|},
\end{align}
where $\mathbf{m}_1$ and $\mathbf{m}_2$ represent any two binary masks. 

The value of IoU ranges from 0 to 1. The closer the value is to 1, the higher the alignment between the model's explanation and the clinical interest, which means that the XAI method has better interpretability in revealing the rationale of its decision-making. 

The AUITC metric provides a comprehensive assessment of alignment by evaluating IoU scores across varying thresholds $\tau$ applied to XAI outputs. It is computed as:
\begin{align}
\text{AUITC} = \int_0^1 \text{IoU}(\mathbf{m}^{\text{gt}}, \mathbf{m}(\tau)) \, d\tau,
\end{align}
where $\mathbf{m}^{\text{gt}}$ is the ground truth clinically significant mask, and $\mathbf{m}(\tau)$ is the binary mask generated by applying a threshold $\tau$ to the XAI output.

\paragraph{SegX Consistently Improves Alignment.} 
Table~\ref{table:SegX} compares the original XAI methods with SegX on the HAM10000 and ChestX-Det10 datasets, using ResNet and DenseNet classification models. The results show that SegX consistently outperforms the original XAI methods in terms of IoU across all datasets and model-XAI combinations. On the HAM10000 dataset, SegX improved both IoU and AUITC for both SHAP and Grad-CAM when applied to ResNet and DenseNet. Similarly, on the ChestX-Det10 dataset, SegX demonstrated smaller but noticeable gains in both metrics, reinforcing its ability to enhance alignment between interpretability and clinically significant regions. 

These findings highlight the effectiveness of the segmentation guidance mechanism in improving the interpretability of the model, ensuring closer alignment between model explanations and clinically relevant regions, thereby increasing the usability of interpretability in medical applications.

\paragraph{SegX Excludes Misleading Explanations.} 
Fig.~\ref{fig:visual comparisons} illustrates interpretability masks generated by the original XAI methods and our SegX approach on the HAM10000 and ChestX-Det10 test sets. The results demonstrate that baseline XAI methods often produce coarser masks that highlight clinically irrelevant regions. In contrast, SegX refines these outputs by focusing more accurately on critical areas of interest, such as lesions or disease-relevant regions.

This refinement, achieved by incorporating clinically significant regions through segmentation, reduces the influence of irrelevant areas in model explanations. By providing clearer and more focused interpretability maps, SegX enhances trust in the model's decision-making process, particularly in scenarios requiring precise identification of clinically significant regions, making it a valuable tool for clinical applications.

\paragraph{SegX's Improvement is Effected by Original XAI.}
Table~\ref{table:SegX} demonstrates that SegX consistently improves alignment between XAI outputs and clinically significant regions on both HAM10000 and ChestX-Det10 datasets. However, the improvements in IoU and AUITC are more pronounced on the HAM10000 dataset. Specifically, SegX achieves a 1-2\% increase in IoU and 2-3\% improvement in AUITC on HAM10000 across both ResNet and DenseNet models. In contrast, the gains on the ChestX-Det10 dataset are relatively smaller, with improvements in IoU and AUITC limited to less than 1\% for most model-XAI combinations.

The greater effectiveness of SegX on the HAM10000 dataset may be influenced by the baseline performance of the original XAI methods. As shown in Table~\ref{table:SegX}, the original XAI methods yield higher IoU and AUITC scores on the HAM10000 dataset compared to the ChestX-Det10 dataset, indicating stronger baseline performance for dermoscopic images. SegX builds upon the outputs of the original XAI methods, and its ability to enhance interpretability relies on the quality of these baseline explanations. On ChestX-Det10, where the original XAI methods perform worse, the improvement achieved by SegX is inherently limited. This suggests that the effectiveness of SegX is related to the initial quality of the XAI outputs it seeks to refine.

\subsection{Results of SegU Evaluation}
In this experiment, we evaluate the effectiveness of SegU in distinguishing between correct and incorrect predictions of the classification model using the IoU-based score $c_{\textbf{IoU}}$ and the AUITC-based score $c_{\textbf{AUITC}}$. By comparing the certainty scores between these groups, we assess whether the SegU approach can effectively differentiate accurate predictions from inaccurate ones.

\paragraph{SegU Effectively Distinguishes Correct and Incorrect Predictions.}
The evaluation results in Table~\ref{table:SegU} shows that SegU achieves consistently higher $c_\text{IoU}$ and $c_\text{AUITC}$ scores for the correct predictions compared to incorrect predictions across all datasets, models, and XAI methods. For instance, on the HAM10000 dataset, Grad-CAM with ResNet achieves a $c_\text{IoU}$ of 0.112 for correct predictions but only 0.015 for incorrect predictions, with a similar trend in $c_\text{AUITC}$ (0.142 vs. 0.022). A similar pattern is observed for the ChestX-Det10 dataset, where SHAP with ResNet achieves a $c_\text{IoU}$ of 0.055 for correct predictions compared to 0.010 for incorrect predictions. This consistent disparity in scores highlights SegU's ability to assess the quality of interpretability outputs relative to model prediction correctness.

These findings suggest that SegU provides a reliable means to quantify the alignment between saliency maps and clinically significant regions, effectively distinguishing correct predictions from incorrect ones. This distinction is critical for clinical decision-making, as it allows clinicians to identify predictions that require additional scrutiny. By quantifying the uncertainty in interpretability outputs, SegU enhances trust in AI-assisted diagnosis and provides an additional layer of validation to improve model reliability in clinical practice.

\paragraph{SegU Struggles with SHAP on Dermoscopic Images.}
The results in Table~\ref{table:SegU} reveal that SegU exhibits only marginal differences in $c_\text{IoU}$ and $c_\text{AUITC}$ scores between correct and incorrect predictions for SHAP on the HAM10000 dataset. For example, the $c_\text{IoU}$ scores are 0.133 and 0.122 for correct and incorrect predictions, respectively, while the $c_\text{AUITC}$ scores are 0.135 and 0.122. These narrow score gaps indicate that SegU struggles to distinguish between the two groups when SHAP is used as the XAI method.

SHAP explanations assume feature independence \cite{aas2021explaining}, resulting in the explanation maps that highlight pixels distributed across both clinically relevant and irrelevant areas. For dermoscopic images, where lesions are typically clear and localised in the centre, this distributed saliency can lead to similar alignment scores for both correct and incorrect predictions, thereby diminishing SegU's ability to distinguish between them effectively.

\section{Conclusion}

In this study, we proposed SegX and SegU to mitigate the limitations of existing KG-XAI methods for medical image analysis, specifically SegX's effectiveness in multi-disease scenarios and SegU's potential to improve the reliability of classification models. Our experiments show that SegX consistently improves the relevance of explanation maps to clinical interests across diverse imaging modalities. Furthermore, to ensure that SegX does not obscure incorrect model predictions, we introduced SegU to measure model reliability and provide clinicians with deeper insights into the trustworthiness of predictions. Both SegX and SegU are model-agnostic, making them versatile tools for enhancing medical decision-making by offering improved explanations and robust uncertainty quantification.

Certain limitations remain while our method demonstrates significant potential. First, the IoU may not be the best measurement to calculate the distance between clinically relevant regions and XAI outputs, as IoU is insensitive to small shifts, shape differences and disproportionately penalties on small regions.  Second, SegX and SegU require the accuracy of the segmentation model, which may not always capture clinically significant regions with sufficient precision. We believe this is not a critical issue thanks to the advancement of (zero-shot) vision foundation models. Thirdly, we validated SegX and SegU on a chest ray diagnosis dataset and a skin cancer recognition dataset. Future work could explore SegX and SegU on diverse imaging modalities, such as MRI and ultrasound, to help validate their generalisability. 

% \section{Supplementary}
% \subsection{Classification Performance}

\bibliography{ref}

\end{document}